# A CONVECTIVE MODEL OF A ROTON

## V.I. Tkachenko[1,2]

[1] *National Science Center "Kharkov Institute of Physics and Technology"*
*The National Academy of Sciences of Ukraine*
*61108, Kharkov 1, Akademicheskaya str., tel./fax 8-057-349-10-78*
[2] *V.N. Karazin Kharkiv National University,*
*61022, Kharkov,4, Svobody sq., tel./fax 8-057-705-14-05*
E-mail: tkachenko@kipt.kharkov.ua

A convective model describing the nature and structure of the roton is proposed. According to the model, the roton is a cylindrical convective cell with free horizontal boundaries. On the basis of the model, the characteristic geometric dimensions of the roton are estimated, and the spatial distribution of the velocity of the helium atoms and the perturbed temperature inside are described. It is assumed that the spatial distribution of rotons has a horizontally multilayer periodic structure, from which follows the quantization of the energy spectrum of rotons. The noted quantization allows us to adequately describe the energy spectrum of rotons. The convective model is quantitatively confirmed by experimental data on the measurement of the density of the normal component of helium II, the scattering of neutrons and light by helium II. The use of a convective model for describing the scattering of light by Helium II made it possible to estimate the dipole moment of the roton, as well as the number of helium atoms participating in the formation of the roton.
**KEY WORDS:** superfluid helium, convection, elementary convective cell, roton, energy spectrum of helium II, density of the normal component of helium II, neutron scattering, light scattering, dipole momentum

## EXPERIMENTS TO DETERMINE THE ENERGY SPECTRUM OF ROTON

In [1] Landau presented a quantitative theory of superfluid Helium, which described almost all known at that time experimental results and predicted the number of new phenomena.

According to this theory, at the temperature below $\lambda$ - point ($T_\lambda = 2,17$ K), Helium, which is called Helium II, at the same time contains two types of thermal excitations: phonons and rotons.

The first type of thermal excitations (phonons) exists in the long-wavelength part of the spectrum and is characterized by a linear dependence of the quasiparticle's energy of its momentum: $\varepsilon(p) = cp$, where $c$ - velocity of ordinary sound in Helium II, which is about 240 $m/s$, $\varepsilon(p)$ and $p$ - the quasiparticle's energy and momentum respectively.

The second type of thermal excitations (rotons) exists in a relatively shorter comparing to phonons wavelength spectrum. At the same time, it is postulated that the relation between energy $\varepsilon(p)$ and momentum $p$ in this part of the energy spectrum is characterized by a parabolic dependence: $\varepsilon(p) = \Delta + (p - p_0)^2 / 2\mu$, where the constants are defined as follows: $\Delta$ - the width of the energy gap between the unexcited and excited levels, $p_0$ - roton's momentum, $\mu$ - effective roton's mass (in units of mass of the atom $^4He$) in which neutrons are scattered.

The postulated dependence of energy on momentum in the short-wave part of the spectrum has been confirmed, according to the authors of a number of publications (see, for example, [2, 3, 4]), in experiments using neutron diffraction analysis when the probe beam of monochromatic neutrons with a specific wavelength is scattered on Helium II. In these experiments the values of the constants of Landau energy spectrum have been identified which are summarized in Table 1:

Table 1: Values of the rotons energy spectrum's constants

| | | | |
|---|---|---|---|
| $\Delta/k = 9.6$, $K$ | $p_0/\hbar = 1.95$, Å$^{-1}$ | $\mu = 1.06$ | [2] |
| $\Delta/k = 8.1 \pm 0.4$, $K$ | $p_0/\hbar = 1.9 \pm 0.03$, Å$^{-1}$ | $\mu = 0.16 \pm 0.02$ | [3] |
| $\Delta/k = 8.65 \pm 0.04$, $K$ | $p_0/\hbar = 1.92 \pm 0.01$, Å$^{-1}$ | $\mu = 0.16 \pm 0.01$ | [4] |

where - $m_{He} = 6.6464836122 \cdot 10^{-24}$ g the mass of the Helium atom [5], $k$ - the Boltzmann constant.

According to [4], the width of the energy gap $\Delta/k$ is not constant, but decreases with increasing temperature by the law $\Delta/k = 8.68 - 0.0084 \cdot T^7$ $K$.

As shown in Table 1, the of the energy spectrum's constants values are different.

This discrepancy points to the need to give the physical sense to the postulated short-wave part of the energy spectrum.

## MODELS OF A ROTON'S STRUCTURE

A number of experimental data about the roton's thermophysical parameters is not yet completed by the establishment of its physical model.

Throughout the Helium II investigations by various methods do not stop attempts trying to describe the roton's structure and properties.

For example, R. Feynman proposed a roton's model as a vortex ring, which consists of six atoms $^4He$ arranged along the ring line with the gaps between them by the order of the Helium atom diameter. Each atom in the ring is rotated in synchronism, being located in the initial position or taking the neighboring, e.g., the left interval. The typical size of a vortex ring is about average atomic distance in a liquid Helium II [6].

Another model of roton's structure [7] is based on the assumption that at temperatures $0.6 \leq T(K) \leq 1.2$ in Helium II exist stable clusters that are bound states of a certain number $N_c$ of atoms $^4He$ ($N_c \gg 1$). The number $N_c$ of atoms in such roton cluster can be determined by the free energy minimum, which is a condition for the stability of a roton. According to [7], the roton cluster must have nearly spherical shape, and the number of Helium atoms - $N_c = 13$. The average radius of the mentioned spherical cluster is estimated at 5.22 Å.

However, in the described roton's models (vortex and spherical) the physical nature describing the forces that hold the Helium atoms in the vortex or in the sphere is missing.

## FREE ENERGY, THE NUMBER OF PHONONS AND PHONON DENSITY'S PART OF THE NORMAL COMPONENT OF HELIUM II

The Bose free energy - for the gas per unit volume is given by [1, 8, 9]:

$$F(T) = -\frac{kT}{(2\pi\hbar)^3} \int_0^\infty \int_0^\pi \int_{-\pi}^\pi \ln(1+n(\varepsilon)) d\vec{p} \tag{1}$$

where $n(\varepsilon(p)) = \left(e^{\frac{\varepsilon(p)}{kT}} - 1\right)^{-1}$ - the Planck distribution function for particles with energy $\varepsilon(p)$, $d\vec{p} = p^2 dp\, do$ - the volume element in momentum space $\vec{p}$, $do$ - the element of solid angle.

For phonons $\varepsilon(p) = cp$, where $c$ - the speed of sound in Helium II, $p$ - phonon momentum modulus.

Integrating by parts in (1), for the phonons free energy gives:

$$F_{ph}(T) = -\frac{kT}{(2\pi\hbar)^3} \int_0^\infty \int_0^\pi \int_{-\pi}^\pi \ln(1+n(cp)) d\vec{p} = -\frac{kT}{(2\pi\hbar)^3} \left[ \int_0^\pi \int_{-\pi}^\pi \ln(1+n(cp)) \frac{p^3}{3} \Big|_0^\infty do + \right.$$

$$\left. + \frac{1}{3kT} \int_0^\infty \int_0^\pi \int_{-\pi}^\pi np \frac{d\varepsilon(p)}{dp} d\vec{p} \right] = -\frac{1}{3(2\pi\hbar)^3} \int_0^\infty \int_0^\pi \int_{-\pi}^\pi npc\, d\vec{p} \tag{2}$$

The first term in square brackets at the upper and lower limits is zero.

The integration of (2) gives the expression:

$$F_{ph}(T) = -\frac{c}{3(2\pi\hbar)^3} \int_0^\infty \int_0^\pi \int_{-\pi}^\pi \left(e^{\frac{cp}{kT}} - 1\right)^{-1} p^3 dp \sin(\vartheta) d\vartheta d\varphi = -\frac{4\pi}{3(2\pi\hbar)^3} kT \left(\frac{kT}{c}\right)^3 \int_0^\infty \frac{x^3}{e^x - 1} dx =$$

$$= -\frac{4\pi^5}{45(2\pi\hbar)^3} kT \left(\frac{kT}{c}\right)^3 = -\frac{E_{ph}}{3} \tag{3}$$

where $E_{ph} = \frac{4\pi^5}{15} kT \left(\frac{kT}{2\pi\hbar c}\right)^3$ - phonons energy in a unit volume of Helium II.

At the same time, the number of phonons per unit volume is determined by:

$$N_{ph}(T) = \frac{1}{(2\pi\hbar)^3} \int_0^\infty \int_0^\pi \int_0^{2\pi} \left(e^{\frac{cp}{kT}} - 1\right)^{-1} p^2 dp \sin(\vartheta) d\vartheta d\varphi = 2\zeta(3) 4\pi \left(\frac{kT}{2\pi\hbar c}\right)^3 \tag{4}$$

where $\zeta(x)$ - Riemann zeta-function of the argument $x$, $\zeta(3) \approx 2.4041$.

Thus, the phonon energy is related to the number of phonons per unit volume of Helium II by the expression

$$E_{ph} = \frac{\pi^4 kT}{36,0617} N_{ph} \tag{5}$$

Phonon gas density is determined from the expression for flow in Helium II. The flow is determined by the momentum in the reference frame moving with the superfluid component:

$$\vec{p} = \vec{j} - \rho\vec{v}_s = \rho_n\vec{v}_n + \rho_s\vec{v}_s - (\rho_n + \rho_s)\vec{v}_s = \rho_n(\vec{v}_n - \vec{v}_s) = \rho_n\vec{w} \tag{6}$$

where $\vec{w} = \vec{v}_n - \vec{v}_s$ - relative velocity of the normal and superfluid components [9].

The same momentum, by definition equals to:

$$\rho_{nph}\vec{w} = \frac{1}{(2\pi\hbar)^3}\int_0^\infty\int_0^\pi\int_0^{2\pi}\left(e^{\frac{cp-\vec{p}\vec{w}}{kT}}-1\right)^{-1}\vec{p}p^2 dp\,sin(\vartheta)d\vartheta d\varphi. \tag{7}$$

For the small difference between velocities $|\vec{w}|$ the Planck distribution function in (7) can be expanded in a series $\vec{p}\vec{w}$ and by retaining only the second term of the expansion (because the first term of the expansion due to asymmetric integrand in symmetrical limits of integration equals zero) one can obtain:

$$\int_{-\infty}^{\infty} n(cp)\,\vec{p}d\vec{p} = \int_{-\infty}^{\infty} n(cp)(p_x\vec{e}_x + p_y\vec{e}_y + p_z\vec{e}_z)dp_x dp_y dp_z = 0. \tag{8}$$

For the calculation of the second term of the expansion (7) let's proceed as follows. Multiply (7) scalar left and right $\vec{w}$. As a result, we have:

$$\rho_{nph}|\vec{w}|^2 = \frac{-1}{(2\pi\hbar)^3}\int_0^\infty\int_0^\pi\int_0^{2\pi}\frac{d}{d(cp)}\left(\left(e^{\frac{cp-\vec{p}\vec{w}}{kT}}-1\right)^{-1}\right)\Bigg|_{\vec{p}\vec{w}=0}(\vec{p}\vec{w})^2 p^2 dp\,sin(\vartheta)d\vartheta d\varphi \tag{9}$$

Product in the right side $\vec{p}\vec{w}$ of (9) in spherical system of momentum is as follows: $\vec{p}\vec{w} = p|w|sin(\vartheta)cos(\varphi)$. Then, after reducing a common factor $|\vec{w}|^2$, the expression (9) takes the form:

$$\begin{aligned}\rho_{nph} &= -\frac{1}{(2\pi\hbar)^3}\frac{1}{c}\int_0^\infty\int_0^\pi\int_0^{2\pi}\frac{dn(cp)}{dp}p^4 sin^3(\vartheta)cos^2(\varphi)dp\,d\vartheta d\varphi = \\ &= -\frac{1}{(2\pi\hbar)^3}\frac{1}{c}\int_0^\infty p^4 d(n(cp))\int_0^\pi sin^3(\vartheta)d\vartheta\int_0^{2\pi}cos^2(\varphi)d\varphi = \\ &= -\frac{1}{(2\pi\hbar)^3}\frac{1}{c}\int_0^\infty p^4 d(n(cp))\frac{\pi^{\frac{1}{2}}\Gamma(2)}{\Gamma\left(\frac{3}{2}+1\right)}\pi = -\frac{1}{(2\pi\hbar)^3}\frac{4\pi}{3}\frac{1}{c}\left(n(cp)p^4\Big|_0^\infty - 4\int_0^\infty n(cp)p^3 dp\right) = \\ &= \frac{4}{3}\frac{E_{ph}}{c^2}.\end{aligned} \tag{10}$$

Thus, the existing ideas about the phonon component of the normal component of liquid Helium II are adequately described by the Planck distribution function for particles with energy $\varepsilon(p)$.

## FREE ENERGY, THE ENERGY AND THE NUMBER OF ROTON DENSITY'S PART OF THE NORMAL COMPONENT OF HELIUM II

The calculation results shown in [1] for the roton density, in which the distribution function of the Planck energy of the roton is represented in the form:

$$\varepsilon_r(\vec{p}) = \Delta + \frac{(\vec{p}-\vec{p}_0)^2}{2\mu} \tag{11}$$

where $\Delta$ - energy roton gap, $\vec{p}_0$ - minimum momentum of $\varepsilon_r(\vec{p})$, $\mu$ - the roton's effective mass.

Expression (11) points to the existence of the directional movement with momentum $\vec{p}_0$ in the normal component of Helium II.

In fact, the entire volume of the liquid Helium II is in equilibrium state, and any aiming motions are absent. This means that in addition to the kinetic energy of a roton in the direction of movement $\vec{p}_0$, equal $\frac{(\vec{p}-\vec{p}_0)^2}{2\mu}$, must be present the kinetic energy of a roton motion in the direction $-\vec{p}_0$, equal $\frac{(\vec{p}+\vec{p}_0)^2}{2\mu}$. In this regard, the roton components energy should be equal to the average energy of the rotons flow directed towards:

$$\varepsilon_r(\vec{p}) = \frac{(\vec{p}-\vec{p}_0)^2}{4\mu} + \frac{(\vec{p}+\vec{p}_0)^2}{4\mu} = \Delta_0 + \frac{p^2}{2\mu} \tag{12}$$

where $\Delta_0 = \frac{p_0^2}{2\mu}$ - roton gap, defined by a photon's momentum.

The number 4 in the denominator in (12) is necessary to ensure the transition to the classical recording of the dependence of energy on momentum in $\vec{p}_0 \to 0$.

The possibility of the existence of a counter-motion of the atoms in Helium II will be discussed below.

Let's consider $\Delta_0 \gg kT$. Then Planck distribution function can be converted to the Boltzmann function. Under these conditions, the free energy per unit volume of the roton component is:

$$F_r(T) = -\frac{kT}{(2\pi\hbar)^3} \int_0^\infty \int_0^\pi \int_{-\pi}^\pi \ln(1+n(\varepsilon_r)) d\vec{p} \approx$$

$$\approx -\frac{kT}{(2\pi\hbar)^3} \int_0^\infty \int_0^\pi \int_{-\pi}^\pi p^2 dp \cdot \sin(\vartheta) d\vartheta d\varphi \cdot \exp\left(-\frac{p^2}{2\mu kT} - \frac{\Delta_0}{kT}\right) = -kT \left(\frac{\mu kT}{2\pi\hbar^2}\right)^{\frac{3}{2}} e^{-\frac{\Delta_0}{kT}}.$$

(13)

Expression (13) was first obtained by Landau [1].

Number of rotons per unit volume is calculated as in (13), and is given by:

$$N_r(T) = \frac{e^{-\frac{\Delta_0}{kT}}}{(2\pi\hbar)^3} \int_0^\infty \int_0^\pi \int_{-\pi}^\pi \exp\left(-\frac{p^2}{2\mu kT}\right) p^2 dp \cdot \sin(\vartheta) d\vartheta d\varphi = \left(\frac{\mu kT}{2\pi\hbar^2}\right)^{\frac{3}{2}} e^{-\frac{\Delta_0}{kT}}$$

(14)

Thus, the free energy per unit volume of the rotons of the Helium II normal component is related to the amount of the roton by the expression $F_r(T) = -kTN_r(T)$.

Roton part of the Helium II normal component's density is determined from the expression (7). Its value is equal to:

$$\rho_{nr}\vec{w} = \frac{e^{-\frac{\Delta_0}{kT}}}{(2\pi\hbar)^3} \int_0^\infty \int_0^\pi \int_0^{2\pi} \vec{p} p^2 dp \sin(\vartheta) d\vartheta d\varphi \cdot \exp\left(-\frac{p^2}{2\mu kT} + \frac{\vec{p}\vec{w}}{kT}\right)$$

(15)

Expanding the exponential in a series of small $|\vec{w}|$, we obtain a simpler expression for the roton part of the normal density:

$$\rho_{nr}\vec{w} \approx \frac{e^{-\frac{\Delta_0}{kT}}}{(2\pi\hbar)^3} \frac{1}{kT} \int_0^\infty \int_0^\pi \int_0^{2\pi} \vec{p}(\vec{p}\vec{w}) p^2 dp \sin(\vartheta) d\vartheta d\varphi \cdot \exp\left(-\frac{p^2}{2\mu kT}\right).$$

(16)

Let's scalar multiply the left and right of (16) on the vector $\vec{w}$. As a result, we have:

$$\rho_{nr} |\vec{w}|^2 \approx \frac{e^{-\frac{\Delta_0}{kT}}}{(2\pi\hbar)^3} \frac{1}{kT} \int_0^\infty \int_0^\pi \int_0^{2\pi} (\vec{p}\vec{w})^2 p^2 dp \sin(\vartheta) d\vartheta d\varphi \cdot \exp\left(-\frac{p^2}{2\mu kT} - \frac{\Delta_0}{kT}\right).$$

(17)

Substitution $\vec{p}\vec{w}$ in a momentum spherical system in a form $\vec{p}\vec{w} = p|w|\sin(\vartheta)\cos(\varphi)$ and subsequent integration of (17) yields $\rho_{nr} = \mu N_r$.

Thus, the number of rotons $N_r$, and with it $\rho_{nr}$, depend on the temperature exponentially.

The above calculations of free energy, energy and the number of quasi-particles per unit volume of the normal component of Helium II match the original calculations for the equilibrium rotons distribution carried by Landau (12) [1].

A little later Landau changed to the non-equilibrium energy spectrum (11), adding at the same time the question about the reason for the non-equilibrium origin.

Thus it can be concluded that the above-noted difference in the experimental determination of the roton energy gap value, the momentum and its effective mass (see. Table 1), and the lack of description of the physical nature of the forces holding Helium atoms in a roton indicates the need of search of the models describing the roton's physical nature and structure.

Therefore, in the present article on the basis of the currently existing sufficiently comprehensive experimental data, a physical model of the roton's origin, defined its dimensions and internal structure, and provides estimates of its thermodynamic parameters.

## ROTONS EQUILIBRIUM ENERGY SPECTRUM

Equilibrium rotons energy spectrum in the form (12) is possible, if it is formed by the volumetric vortex of Helium atoms moving in a cylindrical volume with a diameter $D = D_r$ and height $h = L_z$. The vortex motion of the atoms is arranged so that close to the cylinder axis they moving, for example, along the axis direction (upward), and at the external border of the cylinder - in the opposite direction. Fig. 1 a) shows schematically the vortex motion of Helium II atoms inside a cylindrical cell.

The existence of a cylindrical convection cells with free boundary conditions in a viscous incompressible fluid first studied experimentally and theoretically described in [10]. It is shown that these cells have such geometrical dimensions that the ratio of diameter $D_r$ to cell's height $L_z$ is constant and equal $D_r/L_z = 3.44$.

There are two liquid flows mutually opposite in the vertical section of such cell: at the cell's upper boundary flow velocity is directed out of the center and at the bottom - to the center or vice versa.

It should be noted that the roton's vortex nature was pointed out by Feynman [6, 11]. He wrote about the roton as a vortex ring similar to the ring of smoke. Such vortex momentum is not associated with its translational movement as a whole, but with the movement around the cylindrical surface curved in a closed ring.

Thus roton can be regarded as a cylindrical convection cell. The use of roton energy spectrum of the form (12) is justified, and allows us to consider the Helium II as a whole, as a resting fluid. The spatial distribution of flows in roton refers to its internal structure and only characterizes its intrinsic properties.

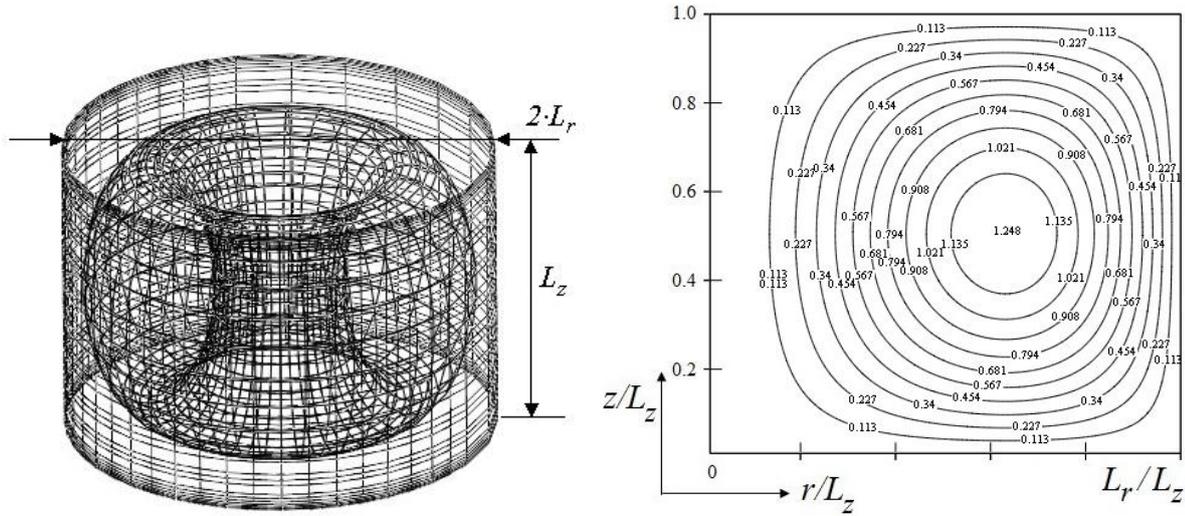

Fig. 1. Schematic representation of the general form of a roton as a cylindrical convection cell - a).
Distribution of Stokes lines $\psi(r, z)$ inside a roton – b)

## ESTIMATION OF THE THERMODYNAMIC PARAMETERS
## OF THE ROTONS FORMATION IN HELIUM II

Let's assume the rotons in Helium II rotons as cylindrical convection cells. Then, to confirm this assumption we should assess the values of the Helium II thermodynamic parameters. These parameters are the coefficients of thermal expansion $\beta$, thermal conductivity $\chi$ and the kinematic viscosity $\nu$ [12]. They specify the value of the Rayleigh number, which, ultimately, will determine the conditions of the origin and existence of a stable cell - the roton.

*Prandtl number.* The Prandtl number $Pr$ is determined by the ratio $Pr = \nu/\chi$. Typically, its value is greater than 1 [13] for the liquid media.

*Rayleigh number.* For the stable existence of convection cells with free boundary conditions Rayleigh number should be equal to the minimal Rayleigh number $R_m = g\beta L_z^3 T_\lambda / \nu\chi = 657.511$ [10, 12, 13].

Let's estimate the values of the dimensional factors that determine the Rayleigh number.

Thermal expansion coefficient for the normal component of superfluid Helium let's define by the expression $\beta(T) = \rho_n^{-1} d\rho_n/dT$. In the temperature range $0.15 \leq T \leq 1.95 \, (K)$ from (10) and (17), $2.58 \leq \beta(T) \leq 25.45 \, (K^{-1})$ and the average value $\bar{\beta} = \dfrac{1}{1.8} \int_{0.15}^{1.95} \beta(x)dx$ is $\bar{\beta} \approx 8.82$.

Let's define dynamic and kinematic viscosity on the basis of data in the scientific literature.

In the review [14] it is noted that the dynamic viscosity coefficient of Helium II is certainly less than $10^{-11}$ Pa·s.

Kapitsa's experiments determined the upper limit of the viscosity of the superfluid component [15]. Assuming laminar flow it was obtained that the value of dynamic viscosity coefficient is $\approx 10^{-11}$ Pa·s.

In [16] the value of the dynamic viscosity coefficient is estimated to be about $10^{-11}$ Pa·s.

In [17] the dynamic viscosity is determined by $\eta = 10^{-14} \left( \dfrac{G \cdot s}{cm^2} \right) = 10^{-14} \left( \dfrac{980 \cdot g \cdot cm \cdot s}{cm^2 s^2} \right) = 0.98 \cdot 10^{-12}$ Pa·s.

Therefore, in further calculations we assume that the coefficient of the dynamic viscosity is $10^{-12}$ Pa·s. Consequently the kinematic viscosity coefficient's order is $\nu \simeq 10^{-12} (\text{Pa} \cdot \text{s}) / \rho \left(\frac{kg}{m^3}\right) \simeq 6,7 \cdot 10^{-15}$ m$^2$ s$^{-1}$.

For the thermal expansion coefficient change range stated above $\beta(T)$, the Prandtl number is in the range $Pr = 3.25...3.83$ [18].

From the above estimates, it follows that, for example, when $\bar{\beta} = 11,2$ the number of Rayleigh $Pr = 3,4988313$ meets the condition of the stable existence of the cylindrical convection cell:

$$R = 3.4988313 \cdot \frac{11.2 \left(\frac{1}{K}\right) \cdot 10 \left(\frac{m}{s^2}\right) \cdot (3.262(m))^3 \cdot 2.17(K)}{(6.7)^2 \left(\frac{m^4}{s^2}\right)} \approx 657.511. \qquad (18)$$

Thus, the above values of the thermodynamic parameters of Helium II correspond to those required to form a stable convection cell.

## THE NATURE AND THE INTERNAL STRUCTURE OF A ROTON

Until now the nature and physical parameters of the rotons are not defined. Therefore, we shall present below the description of the above characteristics of this quasiparticle.

But before describing the nature and the internal structure of the roton, it is useful to trace the process of liquid Helium transition in the superfluid state.

To lower the temperature of liquid Helium below $\lambda$ - the point the permanent pumping of Helium vapor from the cryostat must be performed. Achieving a certain equilibrium temperature located below the $\lambda$ - point is provided by continuously maintaining a certain under pressure in a Helium cryostat [3].

In the process of establishing the thermodynamic equilibrium with falling pressure in the volume of Helium II the origin of the horizontally-layered structure of rotons spatial distribution becomes possible

Lets describe in more detail the formation of the rotons spatial distribution.

### Horizontal multi-layered structure of the rotons spatial distribution

Horizontal multilayered structure of the rotons spatial distribution apparently arises in connection with the formation of the monoatomic superfluid layer when passing through the $\lambda$ - point on the upper boundary of Helium. Under these conditions, apparently, the normal Helium II component located below forms a horizontal transition layer which thickness is sufficiently small and is comparable with the interatomic distance value in Helium II ($L_{0z} \simeq 3,579$ Å). In this transition layer the temperature of the normal Helium II components will increase in the vertical direction from several K degrees at the lower boundary layer (at small deviations of the temperature from $\lambda$ - point) up to zero at the upper boundary layer.

Thus, based on the above-described conditions, we'll came across the problem of Rayleigh-Benard convection in the layer of the viscous, incompressible fluid heated from below [10, 12, 13].

In the above conditions with free boundaries in the transition layer normal Helium II components would occur cylindrical convective cells (read rotons) with a diameter-to-height respect ratio of the order of $D_r/h = 3.44$ [10]. A schematic view of the cylindrical cell and Stokes lines $\psi(r,z) = A \sin\left(\frac{\pi z}{L_z}\right) J_1\left(\frac{\sigma_{1,1} r}{L_r}\right)$ is shown in Fig. 1. In such cell the Helium atoms move along the toroidal surface (Figure 1 a.)) with perturbed velocity $\vec{v}(r,z)$ and its corresponding perturbed temperature $T(r,z)$:

$$v_z(r,z) = A \sin\left(\frac{\pi z}{L_z}\right) J_0\left(\frac{\sigma_{1,1} r}{L_r}\right),$$
$$v_r(r,z) = -A \frac{\pi}{L_z} \frac{L_r}{\sigma_{1,1}} \cos\left(\frac{\pi z}{L_z}\right) J_1\left(\frac{\sigma_{1,1} r}{L_r}\right), \qquad (19)$$
$$T(r,z) = B \sin\left(\frac{\pi z}{L_z}\right) J_0\left(\frac{\sigma_{1,1} r}{L_r}\right),$$

wherein $r, z$ - the horizontal and longitudinal coordinates respectively, $v_z(r,z)$, $v_r(r,z)$ - vertical and horizontal projections of the perturbed speed of the Helium atoms respectively, $A, B$ - constants, $J_k(x)$ - Bessel functions of the first kind $k$ - order of arguments $x$, $\sigma_{1,1} = 3.832$ - the first zero of the Bessel function of the first order ($J_1(\sigma_{1,1}) = 0$).

The toroidal surface inside the cylinder (Fig. 1, a)) corresponds to a certain Stokes line value (Fig. 1 b)) which is defined through the horizontal projection of the perturbed velocity of the substance in convective cell by relation $v_z(r,z) = r^{-1} \partial \psi(r,z)/\partial r$.

From (19) it follows that the horizontal speed has its maximum $r = \sigma_m L_r / \sigma_{1,1}$ on the upper boundary of the cell $z = L_z$, wherein $\sigma_m = 1.8411$ - the argument, in which the function $J_1(x)$ assumes the maximum value $J_1(\sigma_m) = 0.582$.

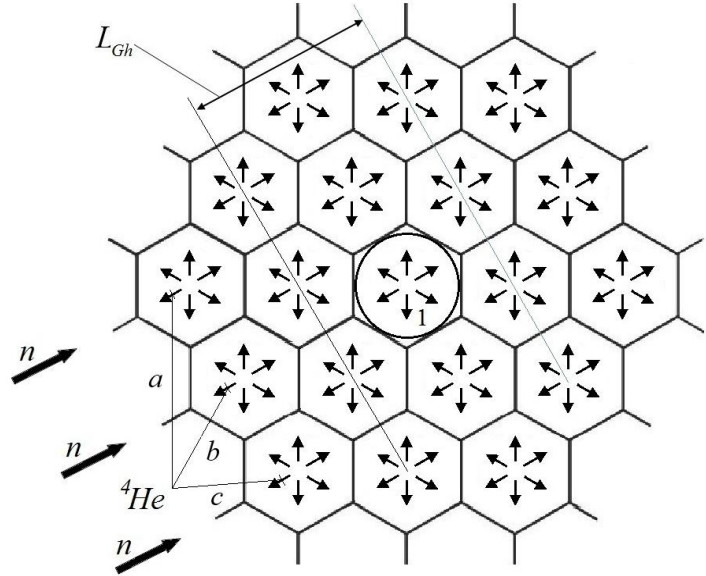

Fig. 2. Horizontal section of the rotons layer on its upper border, indicating the movement direction of the Helium atoms (from the cell center to its outer boundaries).

By increasing the number of cylindrical cells their packaging will occur such that each cell will abut with the six same cells. At the final stage of packaging cylindrical cells fill the entire volume of the transition layer, so that the boundaries between cells will have the form of Benard cells, i.e., hexagons.

The above is shown in Fig. 2. Here numeral 1 is a plan view of the upper boundary of the cylindrical cell in a regular hexagon inscribed. Within each hexagon the from its center arrows indicate schematically the direction of motion of the Helium atoms at the upper boundary of the cell. At the lower boundary of the cell Helium atoms move in the opposite direction. At points $a, b, c$ of counter speed of the Helium atoms has maximum, $L_{Gh}$ - the period of the hexagonal lattice.

Helium II volume filling by the underlying layer of the rotons would occur as follows.

After forming the first layer of convective cells - rotons, below, at the depth of the interatomic distances in Helium II the monomolecular superfluid layer will begin to form again. This movement into deep of the superfluid Helium II can be explained as follows. By lowering the temperature of the Helium II upper boundary the normal flow (warm) of the Helium II components will be directed to the boundary. But according to the second sound theory, normal and superfluid components moving towards each other provide no net flow of the substance [1]. So long cooling of the Helium II upper boundary will contribute to the same duration of the superfluid component penetration in the depth, because time periods of the Helium II components oscillations are same.

Thus at the bottom of the first layer of the convective cells resulting in the above-described moving mechanism of the superfluid component motion the monomolecular layer of the Helium II superfluid components will begin to form. In this case superfluid and normal Helium II components are characterized with the commensurate geometric dimensions.

After the formation of a monomolecular layer of the superfluid component the Helium II normal component located below again forms a horizontal transition layer with a thickness of the interatomic distances in Helium II order. In this new transition layer the second layer of the rotons begins to form. After completion of the rotons second layer formation the superfluid component starts again to accumulate below the second layer and the process of forming the third and subsequent layers of convective cells - rotons is repeated.

Such process of the layered roton gas formation will occur until the thermodynamic equilibrium is achieved and the certain amount of the rotons corresponding given temperature (vapor pressure) is accumulated.

Based on the characteristic size of the elementary convective cell with free boundaries the area of the rotons arrangement in the Helium II normal component is a hexagonal cell with characteristic dimensions: $L_z = 3.262$ Å, $L_r = 1.72 \cdot L_z = 5.61$ Å [10], where - the height of the hexagonal cell $L_z$ selected based on the magnitude of the wave numbers falling on roton minimum.

Thus, from Fig. 2 it follows that a hexagonal crystal lattice with a period $L_z = 3.262$ Å in the vertical direction and $L_{Gh} = 2 \cdot D_r \cdot 2/\sqrt{3} = 25.944$ Å - horizontally is formed in Helium II.

Fig. 2 schematically shows a horizontal section of the upper boundary of the rotons layer. The arrows indicate the movement of the Helium atoms off the cells center to the periphery. Shown in Fig.2 hexagonal structure of the upper boundary of the convective cells pattern repeats itself in depth with the period $L_z$.

The crystal structure of the rotons arrangement described above leads not to the roton-neutron scattering as was previously thought, but the neutron-Helium atoms, which form the rotons, scattering.

In Fig. 2. the lines represent the points of the cell's upper boundary, where the neutrons $n$ are scattered by Helium atoms $^4He$ oncoming neutrons. At these points the Helium atoms have the maximum velocity $-V_{He}$ [10]. At points disposed symmetrically about the center of a cylindrical cell, the maximum velocity of the Helium atoms is equal to $V_{He}$ and is directed along the neutron velocity.

The distance of the noted points from the center of the cell is equal to $R_m = \sigma_{0,1} L_r / \sigma_{1,1}$ where $\sigma_{0,1} \approx 2.405$ and $\sigma_{1,1} \approx 3.832$ - the first zero of the Bessel function of the first kind of orders zero and one respectively [19].

Thus, from the description given above it follows that the cold monoenergetic neutrons are scattered by the periodic lattice $L_z \times L_{Gh}$ consisting of Helium atoms simultaneously moving towards or backwards to the neutrons. Maximum counter velocity of Helium atoms is equal to $-V_{He}$, in the same direction - $V_{He}$.

### THE QUANTIZATION OF THE ROTONS ENERGY SPECTRUM

In the absence of directed flows in Helium II volume (streamless energy spectrum) the rotons energy can be derived in the classical form:

$$\varepsilon_r(p_r, p_z) = \frac{(\vec{p})^2}{2\mu} \qquad (20)$$

Due to the above-noted spatial frequency of the roton gas distribution the rotons energy spectrum (20) should be quantized by (либо in) the vertical $p_z$ and $p_r$ horizontal pulses. Such spectrum quantization (20) can be written as:

$$\varepsilon_{r,nm}(p_r, p_z) = \frac{(p_z - np_{z0})^2 + (p_r - mp_{r0})^2}{2\mu} \qquad (21)$$

where $m = 1, 2, 3\ldots$ and $n = m - 1$ - the quantization numbers, $p_{z0}/\hbar = 2\pi/L_z$ Å$^{-1}$, $p_{r0}/\hbar = 2\pi/L_{Gh}$ Å$^{-1}$.

The actual values of the parameters in the expression (21) can be set on the basis of experimental data on the scattering of neutrons on cold rotons. Let's use the data about the rotons energy spectrum dependence on the wavenumber, which are summarized into a single curve in [20]. Based on this kind of curve we can conclude that the rotons energy spectrum is quantized by the wavenumber and energy.

Fig. 3 shows the energy spectrum of the elementary excitations in Helium II as a function of wavenumber for temperatures below $\lambda$ - point. In this figure, data points from different sources are marked by 1, 2, 3, 4 and 5 [20].

Processing of these data allows the following conclusions regarding the energy spectrum of the quantization parameters rotons (21):

- quantizing the spectrum is observed over the wavenumbers $p_z$ and $p_r$ with constant $p_{z0}/\hbar = 1.925$ Å$^{-1}$ and $p_{r0}/\hbar = 0.242$ Å$^{-1}$, respectively;

- observed in the spectrum quantization of energy with a constant $\Delta_n = ((n+1)p_{r0})^2/2\mu$ where $n = 0, 1, 2$. Such quantization is confirmed by the approximating parabolas $\varepsilon_{r,01}(0, p_z); \varepsilon_{r,12}(0, p_z); \varepsilon_{r,23}(0, p_z)$ which shape describes well the experimental dependence on pulses in a certain range;

Quantitative estimations based on the experimental data presented in Fig. 3, for the quantization number $n = 1$ give following values of the roton's mass and width of the roton's energy gap: $\mu = 0.1649$; $\Delta_1 = 8.606 K$.

It should be noted that the above values of the roton's mass and width of the roton's energy gap correspond to ones shown in Table 1.

Comparison of the experimental data and proposed model indicate the rotons energy spectrum quantization due to the periodic arrangement of rotons in the horizontal and vertical directions in the Helium II volume.

In the proposed model the rotons energy spectrum is represented in a classical form (20). Summing the energies of two rotons derives the energy which is postulated by Landau to justify the appearance of the roton minimum:

$$\bar{\varepsilon}_{r,21}(0, p_z) = \frac{\varepsilon_{r,21}(0, p_z) + \varepsilon_{r,21}(0, -p_z)}{2} = \frac{(p_z - 2p_{z0})^2 + (p_z + 2p_{z0})^2 + 2(p_{r0})^2}{4\mu} = \frac{(p_z)^2 + (2p_{z0})^2 + (p_{r0})^2}{2\mu} =$$

$$= \frac{(p_z)^2}{2\mu} + \bar{\Delta}_{21},$$

where $\bar{\Delta}_{21} = ((2p_{z0})^2 + (p_{r0})^2)/2\mu$ - the Landau's minimum of roton energy.

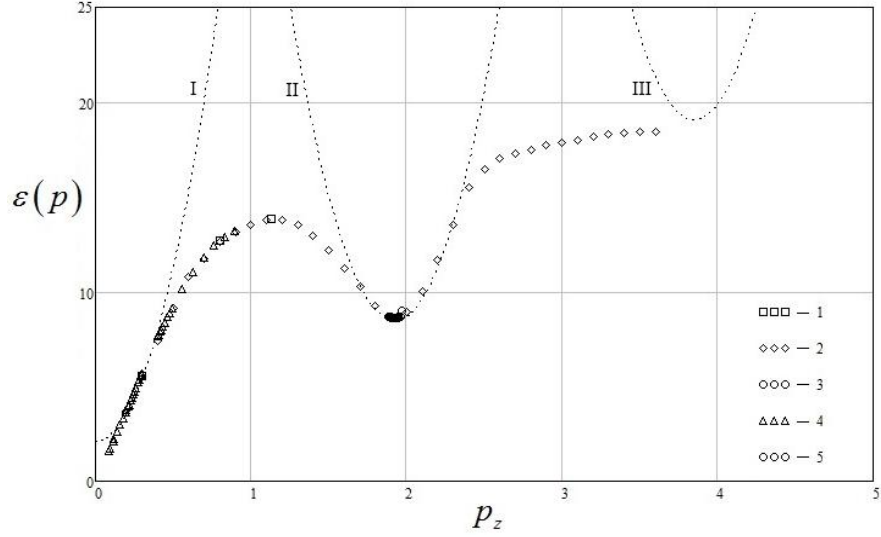

Fig. 3. Energy $\varepsilon$ (in degrees K) of elementary excitations of Helium II as a function of pulse $p_z$ (in Å$^{-1}$) [18]. Parabolas depicted by dots and marked with figures I, II, III correspond to the following quantization numbers:
I - $n=0, m=1$; II - $n=1, m=2$: III - $n=2, m=3$.

Thus, in this section it is shown that the experimental dependence of the normal component Helium II quasiparticles' energy on pulse in the shortwave spectral range can be explained by the streamless character of the rotons energy spectrum and their periodic arrangement in the horizontal and vertical directions.

## THE DENSITY OF THE HELIUM II NORMAL COMPONENT WITH THE STREAMLESS ROTONS ENERGY SPECTRUM

Initially, using the classical representation of the dependence of the rotons energy on a pulse in the form (20) Landau calculated the following thermodynamic quantities of the normal component of helium II: free energy, entropy, heat capacity (per unit mass), and density [8]. Then, in the same work re-published in Physics-Uspekhi, but already with the Appendix, Landau recounted the thermodynamic quantities obtained earlier for the energy of rotons, postulated in the form [1]:

$$\varepsilon_r(p) = \Delta + \frac{(p-p_0)^2}{2\mu} \tag{22}$$

Therefore, there is a natural question about the validity of the application of the roton energy dependence on the pulse in the classical form (20) or in the form (22).

The ratio of the density of the normal component of Helium II to the total density of liquid Helium $\rho_n/\rho$ (by convention, we call this relation by the term "ro-n-to-ro") has the form [1]:

$$\frac{\rho_n}{\rho} = \frac{16}{45}\frac{\pi^5}{c^2}\frac{kT}{\rho}\left(\frac{kT}{2\pi\hbar c}\right)^3 + \frac{\mu N_r}{\rho}. \tag{23}$$

To solve this problem, let's substitute the number of rotons in the expression (23) either in the form $N_r = N_r^{(1)} = \left(\frac{\mu kT}{2\pi\hbar^2}\right)^{\frac{3}{2}} e^{-\frac{\Delta_0}{kT}}$ determined by the energy of the roton in the form (20) (the line $I$ in Fig. 4), or - $N_r = N_r^{(2)} = \frac{2\mu^{\frac{1}{2}} p_0^4}{3(2\pi\hbar^2)^{\frac{3}{2}}(kT)^{\frac{1}{2}}} e^{-\frac{\Delta_0}{kT}}$ determined by the energy of the roton in the form (22) (line $II$ in Fig. 4).

The answer to this question will be a comparative assessment of the coincidence of the theoretical "ro-n-to-ro" dependence on the temperature of the two kinds of roton energy (20) and (22) with experimental points. In this case it is necessary to take into account that the analytical expressions for the temperature dependence of "ro-n-to-ro" are inapplicable for the temperatures near the point $\lambda$ and in a neighborhood of zero [9].

To answer this question, let us use the experimental data on the temperature dependence of "ro-n-to-ro" [20].

In Fig. The 4 markers "×" give the recommended data for experimental measurements of the "ro-n-to-ro" values depending on temperature in the temperature range from 0.15 $K$ to 1.95 $K$.

The curves $I$ and $II$ are constructed from the expression (23) for the number of rotons in a unit volume in the form (20) and (22), respectively:

$$\left(\frac{\rho_n}{\rho}\right)_I = A_I x^4 + B_I x^{1.5} exp\left(-\frac{D_I}{x}\right) \tag{24}$$

$$\left(\frac{\rho_n}{\rho}\right)_{II} = A_{II} x^4 + B_{II} x^{-0.5} exp\left(-\frac{D_{II}}{x}\right) \quad (25)$$

where $x = T/(1\,K)$ - dimensionless temperature; $A_I = 1,19 \cdot 10^{-4}, B_I = 5,66, D_I = 6,7879, A_{II} = 1,21 \cdot 10^{-4}, B_{II} = 204,32,$ $D_{II} = 11,152$ - numerically selected optimal values of the constants in (24), (25).

The optimal parameters were chosen according to the minimum standard deviation of the theoretical dependence on the experimental points.

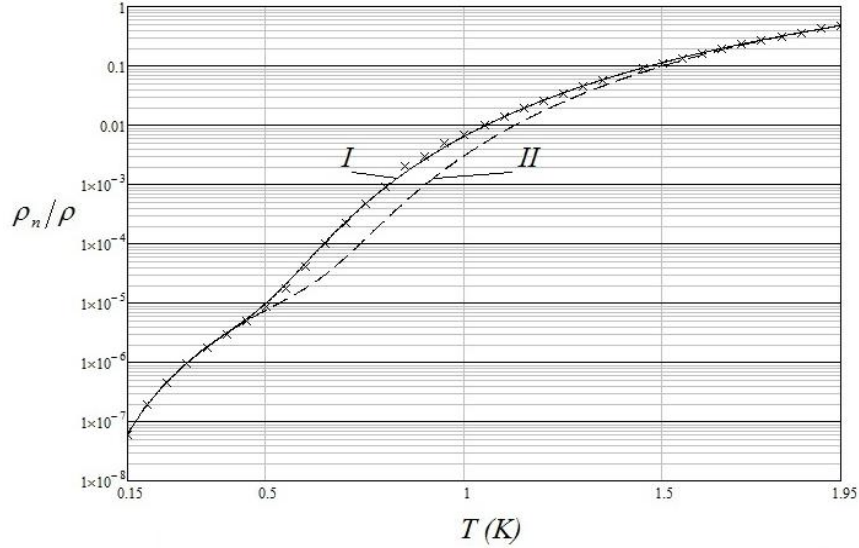

Fig. 4. Dependence of "ro-n-to-ro" on temperature $T$ in the temperature range from 0.15 $K$ to 1.95 $K$. The marks "×" indicate experimentally measured points. The curve $I$ corresponds to the $N_r^{(1)}$, the Curve $II$ - $N_r^{(2)}$.

The smallest standard deviation of the curve $I$ from the experimental points is $1.772 \cdot 10^{-3}$, while for the curve $II$ it is 2.48 times larger – $4.395 \cdot 10^{-3}$.

Thus, in the range of temperatures considered, the "ro-n-to-ro" of Helium II is determined with the greatest degree of accuracy by the streamless rotons energy spectrum in the form (20).

## THE ELASTIC SCATTERING OF NEUTRONS BY MOVING ATOMS OF HELIUM

Unlike the generally accepted concept of the scattering of slow neutrons by Helium II, as inelastic scattering by rotons, we present a different picture of this process. We shall assume that in convective cells the neutrons are elastically scattered by Helium atoms moving towards them. Scattering of neutrons on incidental Helium atoms is not considered as they are slowed down and not detected by the time-of-flight neutron spectrometer [3].

To confirm the validity of this assumption, let us compare the experimental data obtained in the Stockholm experiment on a time-of-flight neutron spectrometer with the results of the theoretical analysis, which are given below.

Let's calculate the neutron scattering parameters for a moving Helium atom.

Let's determine the angle of elastic scattering of a neutron $\psi$ on an atom moving $^4He$ in a cylindrical convective cell. Let the helium atom prior to collision have a maximum horizontal velocity $\vec{V}_{He,\pm} = \mp V_{He} \vec{V}_n / |\vec{V}_n|$ at diametrically opposite points of the upper boundary of the cell (see Fig. 1, 2).

The neutron mass is equal to $m_n = 1,674\,928\,727\,6 \cdot 10^{-24}$ g [21], which is 0.252 times less than the mass of the Helium atom. We assume that the neutron velocity before the collision is $\vec{V}_n$.

As a result of an elastic collision, the neutron will acquire velocity $\vec{V}_{n1}$, and the Helium atom - $\vec{V}_{He1,\pm}$. In the case of an elastic collision, and in the absence of external forces from the laws of conservation of momentum and kinetic energy it is not difficult to obtain the expression for the relative velocity of the scattered neutron $V_{n1}/V_n$:

$$\left(\frac{V_{n1}}{V_n}\right)^2 - 2\frac{V_{n1}}{V_n}\frac{1 \mp \xi_\pm}{1+\zeta}\cos(\psi) + \frac{1 \mp 2\xi_\pm - \zeta}{1+\zeta} = 0 \quad (26)$$

where $\xi_\pm = \zeta \frac{V_{He,\pm}}{V_n}$, $\zeta = \frac{m_{He}}{m_n}$, $\psi$ is the neutron scattering angle.

In equation (26), the upper sign corresponds to the scattering of a neutron by the Helium atom moving in the opposite direction, and in the the same direction for the lower sign.

The solutions of equations (26) have the form:

$$\left(\frac{V_{n1}}{V_n}\right)_{1\pm} = \frac{1\mp\xi_\pm}{1+\zeta}\cos(\psi) + \sqrt{\left(\frac{1\mp\xi_\pm}{1+\zeta}\right)^2 \cos^2(\psi) - \frac{1\mp\xi_\pm - \zeta}{1+\zeta}}, \qquad (27)$$

$$\left(\frac{V_{n1}}{V_n}\right)_{2\pm} = \frac{1\mp\xi_\pm}{1+\zeta}\cos(\psi) - \sqrt{\left(\frac{1\mp\xi_\pm}{1+\zeta}\right)^2 \cos^2(\psi) - \frac{1\mp\xi_\pm - \zeta}{1+\zeta}}. \qquad (28)$$

Calculations show that scattered neutrons with the velocity of Helium atoms have the highest velocity rate $(V_{n1}/V_n)_{1+}$. Therefore, the value of the time shift of the Bragg truncation of the primary beam [3] for a given scattering angle will be minimal, since it is inversely proportional to the velocity of the scattered particle.

This conclusion is based on the following estimates.

We assume that the initial velocity $V_n$ is the same as the neutron velocity after elastic scattering by Vanadium, since it dissipates the neutrons isotropically and without changing the energy [3]. In the Stockholm experiment this velocity is estimated by the value of $V_n = 973.71$ m/s [3].

We set the initial velocity of the neutron in the form $V_n = L/t_0$, where $L$ is the neutron transit distance in the spectrometer, $t_0$ is the transit time.

The neutron velocity after its scattering by Helium II equals to $V_{n1} = L/(t_0 + \Delta t)$, where $\Delta t$ - the magnitude of the time shift of the Bragg truncation of the primary beam. From this equation $(V_{n1}/V_n)_{1+} = 1/(1+\Delta t/t_0)$ we can estimate the relative displacement in time $\Delta t/t_0$:

$$\Delta t/t_0 = \left(\frac{V_{n1}}{V_n}\right)_{1+}^{-1} - 1. \qquad (29)$$

From the expression (29) and the experimental data given in Table 2 [3], it is possible to determine the maximum horizontal counter-flow velocity $V_{He,+}/V_n$ in the convective cell of the scattering center (Helium atom).

Table 2. Maximum horizontal velocity of an atom Helium $V_{He,+}/V_n$ in a convective cell at temperatures $T = 1,4 - 1,5$ °K, depending on the viewing angle $\theta$ [3].

| Type of the filter | The viewing angle, $\theta$ | | $t_0$, μs | $t_0 + \Delta t$, μs | $\Delta t$, μs | $\Delta t/t_0$ *) | $\dfrac{V_{He,+}}{V_n}$ *) | $\left(\dfrac{V_{n1}}{V_n}\right)_{1+}$ *) |
|---|---|---|---|---|---|---|---|---|
| | deg.,° | rad | | | | | | |
| 1 | 2 | 3 | 4 | 5 | 6 | 7 | 8 | 9 |
| Be (○) | 59,4 | 1,037 | 3081 ± 5 | 3415 ± 10 | 334 ± 11 | 0,108 | 0,0355 | 0,902 |
| | 63,6 | 1,110 | 3081 ± 5 | 3395 ± 10 | 314 ± 11 | 0,10191 | 0,065 | 0,908 |
| | 67,7 | 1,182 | 3081 ± 5 | 3347 ± 10 | 266 ± 11 | 0,08634 | 0,1034 | 0,9205 |
| | 69,6 | 1,215 | 3081 ± 5 | 3338 ± 10 | 257 ± 11 | 0,0834 | 0,1136 | 0,923 |
| | 76,3 | 1,332 | 3081 ± 5 | 3305 ± 10 | 224 ± 11 | 0,0727 | 0,145 | 0,9322 |
| | 80,2 | 1,400 | 3081 ± 5 | 3317 ± 10 | 236 ± 11 | 0,0766 | 0,1481 | 0,9289 |
| | 83,0 | 1,449 | 3146 ± 10 | 3392 ± 10 | 246 ± 15 | 0,0782 | 0,1515 | 0,9275 |
| | 86,0 | 1,501 | 3081 ± 5 | 3362 ± 10 | 281 ± 11 | 0,0912 | 0,1431 | 0,9164 |
| | 90,0 | 1,571 | 2992 ± 5 | 3297 ± 10 | 305 ± 11 | 0,1019 | 0,1398 | 0,9075 |
| BeO (×) | 90,0 | 1,571 | 3524 ± 10 | 3917 ± 15 | 393 ± 18 | 0,1115 | 0,1395 | 0,8997 |

*) For the chosen angle $\theta$, the error of the corresponding numbers in columns 7-9 is determined by the largest algebraic sum of column errors 4, 6.

In Fig. 5 shows the experimental points (○ and ×) showing the dependence of the horizontal velocity of the neutron scattering helium atom $V_{He,+}/V_n$ on the viewing angle $\theta$.

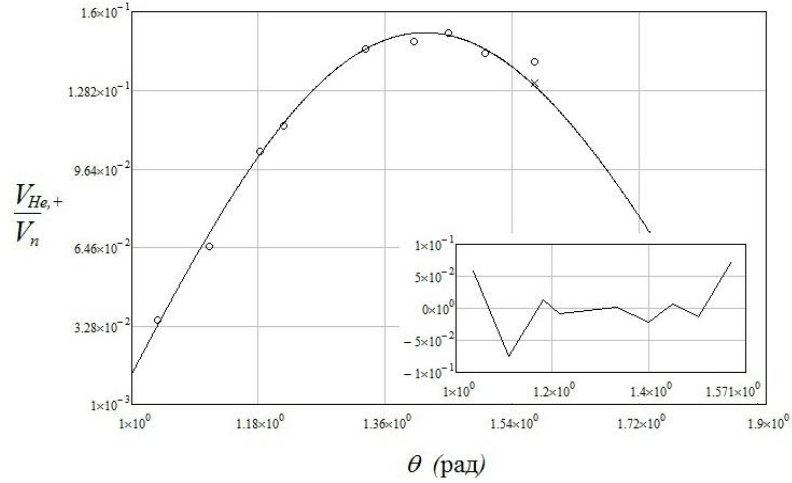

Fig. 5. Dependence of the horizontal velocity of the Helium atom $V_{He,+}/V_n$ in the convective cell on the viewing angle $\theta$ (rad).

From the experimental results presented in Table 2 it follows that in the cell the velocity of the scattering Helium atom depends on the viewing angle $\theta$. At small observation angles, the velocity is small. With the approach of the observation angle to a certain value, in our case about 83 °, the velocity reaches its maximum value, and then decreases again.

The noted above change in the Helium atom's velocity as a function of the viewing angle can be explained on the basis of the dependence of the horizontal velocity of Helium II on the upper boundary of the convective cell on $r$: $v_r(r, L_z) \propto J_1(\sigma_{1,1} r / L_r)$ (19) where $r$ is the distance from the center of the cell to its outer boundary. Such an explanation is possible if we find the dependence of the radius $r$ on the observation angle $\theta$.

To find this dependence, we write down the expression for Archimedes' spiral:

$$\frac{\rho}{\rho_m} = \frac{\varphi}{\varphi_m} \tag{30}$$

where $\rho$ - the polar radius, $\varphi$ - polar angle, $\rho_m = \sqrt{(L_r \sigma_m / \sigma_{1,1})^2 + L_z^2} = 4.2313$ Å, $\varphi_m = arctg(L_r \sigma_m / \sigma_{1,1} / L_z) = 0.6905$.

The type of Archimedes' spiral (30) is determined by the geometric dimensions of the convective cell.

Near the upper boundary of the cell at the point with coordinates $\rho_m, \varphi_m$ (for example, point $b$ in Fig. 2) the horizontal velocity of the Helium atom is maximal. When a neutron collides with a moving Helium atom at this point, the velocity of the scattered neutron $(V_{n1}/V_n)_{1+}$ is the largest of (27), (28). Therefore, further all calculations on the angular distribution of the scattered neutron will be carried out for a fast neutron emitted from a point $\rho_m, \varphi_m$, since only it is fixed by a neutron spectrometer.

Let's proceed to the new variables in (30). We assume $\rho = r + \rho_m$, $\varphi = \theta - \theta_0 + \varphi_m$ that corresponds to the transfer of a point $\rho_m, \varphi_m$ to the pole of polar coordinates, shifted clockwise by a corner $\theta$ to a constant angle $\theta_0$. In this case, the radius $r$ can be represented as:

$$r = \frac{\rho_m}{\varphi_m}(\theta - \theta_0), \tag{31}$$

where the angle's magnitude $\theta_0$ is determined from the experimental data.

The substitution of (31) in (19) near the upper boundary of the cell leads to the expression:

$$v_r(r, L_z) = V_{He,+} = -A \frac{\pi}{L_z} \frac{L_r}{\sigma_{1,1}} J_1\left(\frac{\sigma_{1,1}}{L_r} \frac{\rho_m}{\varphi_m}(\theta - \theta_0)\right) = -A \frac{\pi}{L_z} \frac{L_r}{\sigma_{1,1}} J_1\left(\frac{3.83}{6.51} \frac{4.233}{0.691}(\theta - \theta_0)\right) \tag{32}$$

Thus, the relative horizontal velocity $V_{He,+}/V_n$ will be determined by the equation:

$$\frac{V_{He,+}}{V_n} = -A' J_1(4.1817(\theta - \theta_0)) \tag{33}$$

where $A' = \frac{A}{V_n} \frac{\pi}{L_z} \frac{L_r}{\sigma_{1,1}}$.

Substituting the experimental data of Table 2 in expression (33), and by optimizing the theoretical calculations and experimental data by the method of least mean-square deviation, we determine the values of the constants: $A' = 0.2603$,

$\theta_0 = 0.9755$. The standard deviation is quite small, and is of the order of magnitude $3{,}919 \cdot 10^{-3}$.

The solid line in Fig. 5 shows an optimized curve of the dependence of the relative horizontal velocity of the Helium atom $V_{He,+}/V_n$ on the viewing angle $\theta$. The inset shows the deviation of the theoretical curve from the experimental points. Calculations show that the deviation does not exceed 7.5%.

Thus, in this section it is shown that neutron scattering on a roton can be represented as an elastic collision of a neutron with a Helium atom moving in a cylindrical convective cell. This is indicated by the quantitative agreement of the experimental data obtained earlier by other authors on the scattering of neutrons by Helium II with theoretical calculations of neutron scattering by a helium atom moving in a convective cell.

## SCATTERING OF LIGHT FROM HELIUM II

Along with the neutron diffraction analysis described above, it is possible to study the physical properties of rotons using scattering of light. The data on the measurement of the spectrum, intensity and polarization of light by Raman scattering of argon laser light (514.5 nm wavelength, 1 Watt) by superfluid helium at temperatures in the interval 1.16 $K$ and 2.14 $K$ are given in [22].

In experiments, the incident linearly polarized laser beam and scattered light are located in a horizontal plane. The scattered light was collected at an angle of 90° to the incident radiation within a solid angle of about 0.08 sr. To ensure the maximum level of the detected signal, the vector of the electric field of the incident laser radiation was oriented in the horizontal direction [23]. Analysis of the intensity of the scattered radiation spectrum clearly showed the presence of an asymmetric sharp peak shifted by an energy of 18.5 ± 0.5 $K$ relative to the energy of the incident radiation.

The following work on the Raman scattering of laser radiation by superfluid helium [24] used the same experimental setup as in the original paper [22]. However, instead of the diffraction monochromator, a Fabry-Perot spectrometer was used, which free spectral range was 48.6 $K$, which roughly corresponds to the triple shift of two-roton Raman scattered light. The light source was an argon ion laser (wavelength 488.0 nm). It was shown in this paper that the shift of the energy of the scattered radiation relative to the energy of the incident radiation is 17.022 ± 0.027 $K$ at a temperature of 1.2 $K$.

The measurements carried out in [1 - 3] have shown that the investigation of light scattering is more accurate, in comparison with the neutron diffraction tool, for describing elementary excitations in superfluid Helium. The very high resolution of this method makes it possible to accurately measure such characteristics of excitations in liquid helium, as energy and lifetimes of the roton. Let's describe the above results of experiments on Raman scattering of light on the basis of the convective model of the structure of the roton proposed in this paper.

### The dipole moment of a roton

Let's define the number of the Helium atoms which are involved in the convective motion in a roton.
It follows from the calculations that there are 7 Helium atoms in the volume of one roton:

$$N_{Rot} = \frac{\rho}{m_{He}} V_{Rot} = \frac{0.145 \frac{g}{cm^3}}{6.646 \cdot 10^{-24} g} 3.14 \cdot (5.61)^2 \cdot 3.262 \cdot 10^{-24} = 7.095 \tag{34}$$

where $\rho = 0.145$ g/cm³ [20] - the Helium II density, $V_{Rot} = \pi L_r^2 L_z$ - the roton's volume.

To determine the dipole moment of the roton, it is necessary to calculate the dipole moment of the accelerated atom, using the expression [25]:

$$\vec{d}(r,z) = \gamma \dot{\vec{v}}(r,z) \tag{35}$$

where $\dot{\vec{v}}(r,z)$ is the acceleration of the helium atom, $\gamma = \frac{m_{He} \alpha_0}{2Z|e|}$, $\alpha_0 = 0.21 \cdot 10^{-24}$ cm³ is the polarizability of the helium atom [26], $m_{He}$ and $Z = 2$ is the mass and charge of the nucleus of the Helium atom, $e$ is the elementary charge. In (35) the parameter $\gamma$ has the value $\gamma \approx 7.269 \cdot 10^{-40}$ g cm³/u. charges the CGS.

From (35) and (19) we define the inertial dipole moment of a helium atom moving along a closed trajectory. Based on the axial symmetry of the inner structure of the roton, its dipole moment should be oriented in the vertical direction. Let's show this.

We assume that in the vertical plane passing through the cell axis, two helium atoms move along at closed and symmetric trajectories with respect to the cell axis. Let's determine the projections of their total dipole moments:

$$d_{rs}(r,z) = d_r(r,z) + d_r(-r,z) = \gamma \left[ \dot{v}_r(r,z) + \dot{v}_r(-r,z) \right],$$
$$d_{zs}(r,z) = d_z(r,z) + d_z(-r,z) = \gamma \left[ \dot{v}_z(r,z) + \dot{v}_z(-r,z) \right], \tag{36}$$

where $\dot{v}_r(r,z) = v_r(r,z) \frac{\partial v_r(r,z)}{\partial r} + v_z(r,z) \frac{\partial v_r(r,z)}{\partial z}$, $\dot{v}_z(r,z) = v_r(r,z) \frac{\partial v_z(r,z)}{\partial r} + v_z(r,z) \frac{\partial v_z(r,z)}{\partial z}$.

Taking into account the dependence of the velocity projections $v_r(r,z)$ and $v_z(r,z)$ on the coordinates $r$ and $z$ (19), it is not difficult to show that the horizontal dipole moment of the roton is equal to zero $d_{rs}(r,z) = 0$ and the vertical one is different from zero $d_{zs}(r,z) \neq 0$.

Starting from (36), we can state that two helium atoms moving at closed and symmetric trajectories relative to the cell axis in the same plane have the bound state.

The total dipole moment of two bound Helium atoms is:

$$d_{zs}(r,z) = \pi \frac{\gamma A^2}{L_z}\left[J_0^2\left(\frac{\sigma_{1,1}r}{L_r}\right) + J_1^2\left(\frac{\sigma_{1,1}r}{L_r}\right)\right]\sin\left(2\pi\frac{z}{L_z}\right), \tag{37}$$

and represents two differently directed dipoles with a cylindrically symmetric distribution of the dipole moment in space.

To estimate the magnitude of the total dipole moment from (33) we determine the amplitude of the velocity $A$: $A = 179.71$ m/s.

Then the maximum value of the total dipole moment for bound Helium atoms is reached on the roton axis, and has the order:

$$d_1 = \left|d_{zs}\left(0, \frac{L_z}{4}\right)\right| = 0.226 \cdot 10^{-4} \text{ D}. \tag{38}$$

Hence we can assume that in the convective motion in the roton three pairs of bound Helium atoms (37) are involved which are located in three planes shifted along the azimuth by 120°.

Proceeding from the foregoing, let us estimate the maximum total dipole moment of the roton in the vertical direction $D_{max} = 3d_1 = 0.678 \cdot 10^{-4}$ D. In the horizontal direction, without taking into account the seventh Helium atom, the roton's dipole moment equals to zero.

Thus, the estimates given above show that the roton's dipole moment consists of two dipoles oriented in the vertical direction and directed in opposite directions (see (37)). The magnitude of the dipole moment on the roton axis is of the order of magnitude of the experimentally measured [27].

### The seventh Helium atom

From the condition of Helium II density constancy it follows that the seventh Helium atom is located inside the cell and moves with thermal velocity without leaving its limits. For the long confinement of the seventh Helium atom in the roton and taking into account the problem's symmetry, it is necessary to require that it moves along a circular orbit about the axis of the cell in the plane $z = L_z/2$. In this case, the presence of the seventh Helium atom oscillating inside the roton will lead to the fact that the total horizontal dipole moment of the roton $d_\Sigma = d_{rs} + d_{r7}$ will be different from zero:

$$d_\Sigma = d_{rs} + d_{r7} = \gamma \dot{v}_{r7}. \tag{39}$$

In (39) the acceleration of the Helium atom is determined by the expression $\dot{v}_{r7} = -r\omega_0^2 cos(\omega_0 t) sin(\pi z/L_z)$, and is given by the dependence of the oscillator radius on the time in the form: $R_{os} = r cos(\omega_0 t) sin(\pi z/L_z)$, where $0 \le r \le 5.61$ Å is the radius of Helium atom's motion inside the roton, $\omega_0$ - the frequency of the oscillations, and $t$ - the time.

The assumption made above about the motion of the helium atom along a circular orbit can be physically justified by the action of Van der Waals orientation forces [28], which characterize the interaction between the electric dipoles of the roton and the seventh Helium atom.

The total energy of the orientational interaction of dipoles is determined by the sum of the interaction energy of the Helium atom dipole with two differently directed dipoles of the roton [28]:

$$U(r) = -2\gamma R^{-3}|d_{zr}||d_{r7}|cos(\vartheta_d), \tag{40}$$

where $R$ - the distance between the dipoles centers, $\vartheta_d$ - is the angle between the directions of the roton dipoles and the dipole of the Helium atom. The coordinates of the centers of the dipoles are given in the form: for the seventh helium atom – $(r, L_z/2)$; for two roton dipoles - $(0, (2\pm 1)L_z/4)$.

Proceeding from the assumption made above about the long confinement of the seventh Helium atom inside the roton, it can be stated that the distance between the centers of the dipoles is constant. Therefore, assuming $R = R_0 = const_1$ and $\vartheta_d = \vartheta_{d0} = const_2$ we determine from (40) the coordinate of the minimum energy of the dipoles orientational interaction.

Fig. 6 shows the dependence of the energy of the dipoles orientational interactiones of the roton and the atom in relative units on the distance $r/L_r$ from the roton axis.

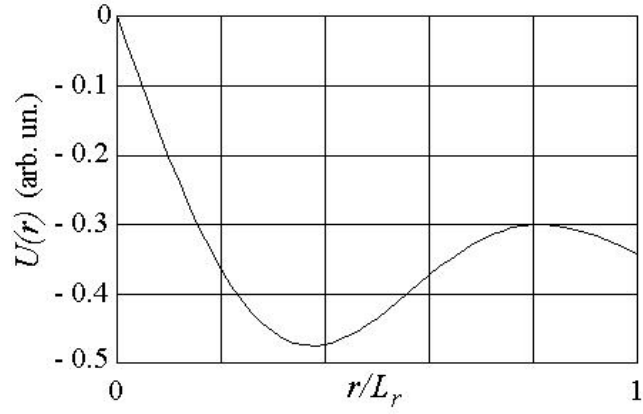

Fig. 6. Dependence of the dipoles orientational interaction energy of the roton and Helium atom in relative units from a distance from the axis of roton.

From the Fig. 6 it follows that the minimum energy of the dipoles orientational interaction is observed at a distance of $r = 0.38 L_r$.

From this follows the estimate of the radius of the circular orbit of the seventh helium atom: $r = 2.13$ Å.

### Raman scattering on the Helium II dipoles

Let's consider the phenomenological model of Raman scattering of light by the dipole moments of rotons in Helium II. According to the proposed model of the roton's structure, the Helium atoms accelerated inside it are polarized under the action of inertial forces, which can be brought into line with the strength of the effective electric field $\vec{E}(r,z)$. The vector of electrical induction in a medium is determined by the electric field strength and the polarization of the medium: $\vec{D}(r,z) = \vec{E}(r,z) + 4\pi \vec{P}(r,z)$. In the absence of external charges and currents, when $\text{div}(\vec{D}) = 0$ and $\frac{\partial \vec{D}}{\partial t} = 0$, the electric induction vector equals to zero. Then the electric field strength in the roton is determined by the expression:

$$\vec{E}(r,z) = -4\pi \vec{P}(r,z), \tag{41}$$

where the polarization of the medium is determined by the expression $\vec{P}(r,z) = 3 d_{zs}(r,z) \vec{e}_z + d_{r7} \vec{e}_r$, and $\vec{e}_z$, $\vec{e}_r$ are the unit vectors along the axis and the roton radius, respectively.

The horizontal component of the polarization of the medium takes part in the scattering of polarized laser light with the electric field vector oriented in the direction of observation $P_x(0, L_z/2) = d_{r7} = -\gamma r \omega_0^2 \cos(\omega_0 t) \sin(\pi z / L_z)$.

We represent the time part of the incident laser light wave $\Omega$ in the form $E(t) = E_0 \cos(\Omega t)$. In the field of this radiation the polarizability and hence the parameter $\gamma$ can be represented as an expansion in the amplitudes of small perturbations $q$ [29]:

$$\gamma(q) \simeq \gamma(0) + \left.\frac{d\gamma(q)}{dq}\right|_{q=0} q, \tag{42}$$

where $q = E_0 \cos(\Omega t)$.

Then the strength of the electric field of the scattered laser radiation takes the form:

$$E_x\left(\frac{L_z}{2}, q\right) = 4\pi r \omega_0^2 E_0 \left( \gamma(0) \cos(\omega_0 t) + \gamma_1\left(\frac{L_z}{2}, 0\right) \left( \cos((\Omega - \omega_0)t) + \cos((\Omega + \omega_0)t) \right) \right), \tag{43}$$

where $\gamma_1\left(\frac{L_z}{2}, 0\right) = \frac{1}{2} \left.\frac{d\gamma(q)}{dq}\right|_{q=0}$.

As one can see, the strength of the electric field of the scattered light (43) has the radiation with a frequency $\Omega$ shifted on a frequency $\omega_0$. Starting from (39) and experimental data [3], we determine the radius of motion of the seventh Helium atom at which the frequency shift is observed on the value $\omega_0 = 17\,K = 35.4229 \cdot 10^{10}$ Hz:

$r\omega_0 = \sqrt{\dfrac{2kT}{m_{He}}} = \sqrt{\dfrac{2 \cdot 1.38 \cdot 10^{-16} \dfrac{erg}{K} \cdot 1.2\,K}{6.646 \cdot 10^{-24}\,g}} = 7059.4\,\dfrac{cm}{s}$. Hence the radius of the seventh Helium atom motion is

$r = R_{He} = \dfrac{7059.4}{35.4229} 10^{-10} cm = 1,993$ Å. The obtained radius value quantitatively corresponds to that given in the previous section: $r = 2.13$ Å.

Thus, the laser radiation scattering on a roton with the parameters described above corresponds quantitatively to the experimental data on Raman scattering of light on Helium II. Such correspondence indicates the possibility of using the proposed model of the roton's nature and structure for describing the thermophysical properties of Helium II.

## CONCLUSION

An analysis of the currently existing representstions of such thermodynamic characteristics of Helium II phonons and rotons as free energy, energy, quantity and also the corresponding part of the density of the normal component of Helium II is presented in the article. At present the thermodynamic parameters of rotons (free energy, energy, number of rotons, roton part of the density of the normal component of Helium II) have an indeterminate value: either the roton energy in the Planck distribution function is minimal in the vicinity of a certain momentum value $p_0 \neq 0$, or it is minimal at zero momentum $p_0 = 0$. Physically this means that either in Helium II the equilibrium energy spectrum is characterized by a directed flux of rotons, or the equilibrium energy spectrum of rotons is streamless.

The paper proposes to consider the streamless energy spectrum for describing the roton part of the normal component of Helium II and a physical substantiation of this approach is given. This proposal is based on the involvement in the description of physical processes in Helium II of such process as convection, which, until now, has not been considered in problems on superfluidity. Estimated calculations show that the thermodynamic parameters of Helium II (the coefficients of thermal expansion, kinematic viscosity and thermal diffusivity) correspond to those at which there are possible to form the stable cylindrical convective cells such as Benard cells.

An analysis of the process of obtaining superfluid Helium from liquid Helium shows that passing through $\lambda$ - point, the convective processes in the transition layer of the normal component of Helium II form horizontal layers of thickness of the order $h = 3.262$ Å, in which rectangular hexagonal prisms with the same height and edge length $c = 3.243$ Å are densely packed. In such hexagonal prisms the cylindrical convective cells of height $h = L_z = 3.262$ Å and diameter $D_r = 5.61$ Å are inscribed, which represent the rotons. For such convective cells the spatial distribution of the convective velocity of Helium atoms is described. Vertical and horizontal spatial periods of the distribution of the roton gas in the volume of Helium II are determined. It is shown that due to the spatial periodicity of the rotons distribution, their energy spectrum is quantized along the vertical and horizontal momentum (wave number). Such spectrum quantization makes it possible to describe with sufficient accuracy the experimental dependence of the energy spectrum of the quasiparticles of the normal component of Helium II on momentum (wave number). It is shown that the theoretically calculated $\rho_n/\rho$ curve with a streamless ($p_0 = 0$) energy spectrum in the wide temperature range (from 0.15 $K$ to 1.95 $K$) closely matches the experimental data. For the streamless energy spectrum it has been theoretically shown and experimentally confirmed that the scattering of slow neutrons occurs not on rotons, but on Helium atoms moving in them. The use of a convective model for describing the scattering of light by Helium II made it possible to determine the dipole moment of the roton as well as the number of Helium atoms participating in the formation of the roton. It is shown that the value of the calculated dipole moment of the roton coincides in order of magnitude with the experimentally measured value. Estimates show that in average seven Helium atoms are involved in the formation of the roton. It is shown that two Helium atoms in the roton are in a bound state i.e. synchronously move with convective velocity along closed trajectories in the vertical plane passing through the roton axis. There are three such bound pairs and they move in three planes located at equal angles in the azimuth. It is estimated that the seventh Helium atom is located inside the roton. It moves in a horizontal plane at a half-height of the roton along a circular trajectory. Based on the analysis of the energy of the dipoles orientational interaction a theoretical estimate is given. The obtained estimate of the radius of motion of the seventh Helium atom with a reasonable degree of accuracy corresponds to the experimental data on the light scattering by Helium II. In conclusion, a statement is drawn on the applicability of the convective model of the roton's nature and structure for the description of the Helium II termophysical properties.